\documentclass[aps,preprint]{revtex4}%
\usepackage{amsfonts}
\usepackage{amsmath}
\usepackage{amssymb}
\usepackage{graphicx}%
\providecommand{\U}[1]{\protect\rule{.1in}{.1in}}

\begin{document}
\title{Chiral tunneling through time-periodic potential in graphene}
\preprint{ }
\author{M. Ahsan Zeb}
\affiliation{National Centre for Physics, Islamabad 45320, Pakistan}
\affiliation{Department of Physics, Quaid-i-Azam University, Islamabad 45320, Pakistan}
\author{K. Sabeeh$^{\dagger}$}
\affiliation{Department of Physics,Quaid-i-Azam University, Islamabad 45320, Pakistan}
\author{M. Tahir}
\affiliation{Department of Physics, University of Sargodha, Sargodha, Pakistan}
\pacs{PACS number}

\begin{abstract}
Chiral tunneling through a harmonically driven potential barrier in graphene
monolayer is considered in this work. Since the quasiparticles in this system
are chiral in nature, tunneling is highly anisotropic, we determine the
transmission probabilities for the central and sidebands as the incident angle
of the electron beam is changed . Furthermore, we investigate how the
transmission probabilities change as the width, amplitude and frequency of the
oscillating barrier is changed. An interesting result of our study is that
perfect transmission for normal incidence that has been reported for a static
barrier persists for the oscillating barrier, manifestation of Klein tunneling
in a time harmonic potential.

\end{abstract}
\volumeyear{year}
\volumenumber{number}
\issuenumber{number}
\eid{identifier}
\date[Date text]{date}
\received[Received text]{date}

\revised[Revised text]{date}

\accepted[Accepted text]{date}

\published[Published text]{date}

\startpage{1}
\endpage{2}
\maketitle

\section{\textbf{INTRODUCTION }}

Advancement in technology has led to active investigation of electron
transport in semiconductor nanostructures in time-dependent fields. The
additional degree of freedom provided by the time dependence has led to the
appearance of new phenomena in electron transport, for a review see \cite{c15}
and references therein. Engineering of the confinement potential and band
structure has allowed the possibility of studying photon assisted tunneling
(PAT), where inelastic tunneling events occur in the presence of an ac field,
in various driven systems. This topic is not only of academic interest but
also has device applications. Early studies of PAT include the work of Dayem
and Martin who provided evidence of absorption and emission of photons in
tunneling transport in experiments on superconducting films in the presence of
microwave fields \cite{c13}. Soon after this, Tien and Gordon theoretically
justified this observation \cite{c14}. They assumed a time harmonic potential
difference produced between the two films by a microwave field and solved the
time-dependent Schrodinger equation for the system. Their photon assisted
transport model accounted for transmission in the side bands in the presence
of microwave radiation.The basic physical idea behind photon assisted
tunneling is that an oscillating potential can lead to in-elastic tunneling
where the electrons exchange energy quanta (photons) with the oscillating
field. In such systems, a harmonically driven in time potential results in
exchange of energy with electrons in the units of modulation quanta
$\hbar\omega,$ $\omega$\ being the modulation frequency. Therefore, electrons
at energy $E$ can be transferred to the sidebands at energies $E\pm
n\hbar\omega$ $(n=0,\pm1,\pm2,..)$\ while traversing a region of space
subjected to such a time-harmonic potential. The prototypical tunneling
structure, which is an essential element of nanostructures where electron
tunneling is investigated, is a single barrier. A common model in these
studies is a time-modulated potential that has a finite spatial profile.
Standard electron transport through various types of time-oscillating
potential regions has been studied previously. More pertinent to the work
undertaken here is that of Buttiker and Landauer. They investigated the
traversal time of particles interacting with a barrier with time-oscillating
height \cite{c16a,c16}. Furthermore, M. Wagner wrote a series of papers on
photon assisted transport through quantum wells and barriers with oscillating
potentials \cite{c17}. Exchange of photons between the oscillating potential
and electrons transfers electrons to the sidebands with a finite probability.
Wagner determined these transmission probabilities using transfer matrix
methods and discussed them as a function of the dimensionless parameter
$\alpha$ which is the ratio of the amplitude of the time oscillating potential
to its modulation energy. There are other contributions to this field that are
relevant to our work and these have been put together in\cite{c18}. Recently,
single layer carbon crystals (graphene monolayer) were fabricated which has
generated considerable interest in finding a material that can replace silicon
in micro-electronic devices. The idea of carbon based nanoelectronics has been
around since the discovery of carbon nanotubes. The recent fabrication of
graphene monolayer has provided another avenue for carbon based electronics.
Devices based on photon-assisted electron tunneling require the consideration
of electron transport in time-harmonic potentials. For graphene based PAT
devices it is essential to consider transport of charge carriers in graphene
through time-harmonic potentials. To this end, we undertake the study
presented here realizing that quasiparticles in graphene systems are quite
different from the standard electrons that we encounter in conventional
semiconductor based heterostructures. At low energies, quasiparticles
(electrons and holes ) in graphene are described by the relativistic
Dirac-like equation and possess charge conjugation symmetry as a single
equation describes both particles (electrons) and antiparticles (holes). This
is due to the crystal structure of graphene which is a layer of carbon atoms
tightly packed in honeycomb lattice. It can be thought of as the superposition
of two equivalent triangular sublattices conventionally called sublattice A
and B. Quantum mechanical hopping between these sublattices results in the
formation of two cosine-like energy bands. Intersection of these bands near
the edges of Brillouin zone (Dirac points) leads to the conical energy
spectrum $E=\pm\hbar v_{F}k$ (with the effective Fermi speed ( $v_{F}%
=10^{6}m/s).$ Above zero energy, the charge carriers in these systems are
electrons which are usually termed Dirac electrons. The 2D Dirac-like spectrum
was confirmed recently by cyclotron resonance measurements and also by angle
resolved photoelectron spectroscopy (ARPES) measurements\cite{3}. Recent
theoretical work on graphene multilayers has also shown the existence of Dirac
electrons with a linear energy spectrum in monolayer graphene\cite{4}. The
Dirac equation implies that the quasiparticles in graphene are chiral,
tunneling through potential barriers in these systems is significantly
different from systems where tunneling of standard electron occurs such as the
two-dimensional electron gas (2DEG) systems realized in semiconductor
heterostructures. Chiral nature of particles in graphene results in quantum
tunneling being highly anisotropic where relativistic effects such as perfect
transmission through high and wide barriers can occur (Klein
tunneling\cite{klein})\cite{c5}. This occurs due to the conservation of
chirality in interaction with the barrier, electrons in graphene can propagate
to hole states through a high barrier without any damping. The study of this
effect is relevant to the development of future graphene based devices. From a
basic research point of view, graphene based systems, due to their lower
`light speed', can be quite useful for studying relativistic effects.
Moreover, the role of chirality can be highlighted in electron transport in
graphene. In graphene-based systems, electronic transport through barrier
structures has been recently investigated
\cite{c5,c7,c11,c23,c24,c25,c26,c27,c28}. In this work, we consider the
transport of Dirac electrons in monolayer graphene through a harmonically
oscillating in time square potential barrier. When standard electrons pass
through a region which is subjected to time harmonic potentials, electronic
transitions\ from central band to sidebands occur. Here, when transmission of
Dirac electrons is considered, we also find transitions from the central to
sidebands at energies $E\pm n\hbar\omega$ $(n=0,\pm1,\pm2,..)$ and determine
the transmission probabilities for the sidebands. Moreover, we investigate how
the transmission probabilities change as various perimeters involved in the
problem are varied with emphasis on the chiral nature of tunneling.

\section{\ FORMULATION}

We consider monolayer graphene sheet in the $xy$-plane. The square potential
barrier is taken to be in the $x$-direction while particles are free in the
$y$-direction. Width of the barrier is $a$, height of the barrier is
oscillating sinusoidally around $V$ with amplitude $V_{1}$ and frequency
$\omega$. Electrons with energy $E$ are incident from one side of the barrier
in monolayer making an angle $\phi_{0\text{ }}$ with the $x$-axis and leave
the barrier with energy $E\pm n\hbar\omega$ $(n=0,\pm1,..)$ making angles
$\phi_{n\text{ }}$ after transmission and $\pi-\phi_{n\text{ }}$ after reflection.

Let us consider the Hamiltonian $H$ describing the system%
\begin{equation}
H_{\text{ }}=H_{0\text{ }}+H_{1\text{ }} \label{1a}%
\end{equation}
where $H_{0\text{ }}$ is the Hamiltonian for the static case where the barrier
height is not changing with time and $H_{1\text{ }}$describes the harmonic
time dependence of barrier height, given by%
\begin{align}
H_{0\text{ }}  &  =-i\hbar v_{F}\sigma.\nabla+V\label{1b}\\
H_{1\text{ \ }}  &  =V_{1}Cos(\omega t)
\end{align}
$V,$ $V_{1}$ are the static square potential barrier and the amplitude of the
oscillating potential, respectively. Both $V$ and $V_{1}$ are constants for
$0\leq x\leq a$ with $a$ positive and are zero elsewhere.\ $\sigma=(\sigma
_{x},\sigma_{y})$ are the Pauli matrices, $v_{F}$ is the Fermi velocity.

Solutions of the Dirac equation in the absence of the oscillating potential,
$H_{0\text{ }}\Psi=E\Psi,$ are given in \cite{c5} and can be used for
constructing solutions to the time-dependent problem. For the tunneling
problem, we consider the incoming electrons to be in plane wave states
$\Psi_{i\text{ }}(x,y,t)$ at energy $E$%
\begin{equation}
\Psi_{i\text{ }}(x,y,t)=e^{ik_{y}y}\binom{1}{s_{0\text{ }}e^{i\phi_{0}}%
}e^{ik_{1}^{0}x}e^{-iEt/\hbar} \label{7}%
\end{equation}
where $k_{1}^{0}$ and $k_{y}$ are the $x-$ and $y-$component of the electron
wavevector, respectively. $s_{0}=sgn(E)$ and $\phi_{0}$ is the angle that
incident electrons make with the $x$-axis.

Reflected and transmitted waves have components at all energies $E\pm
l\hbar\omega$ $(l=0,\pm1,..)$ since the oscillating potential barrier can give
and take energy away from electrons in units of $\hbar\omega$. This change in
energy causes only the $x$-component of momentum to change. Hence,
wavefunctions $\Psi_{r\text{ }}(x,y,t)$ for reflected and $\Psi_{t\text{ }%
}(x,y,t)$ for transmitted electrons, respectively are%
\begin{equation}
\Psi_{r\text{ }}(x,y,t)=e^{ik_{y}y}%
{\displaystyle\sum\limits_{l=-\infty}^{l=\infty}}
r_{l}\binom{1}{-s_{l\text{ }}e^{-i\phi_{l}}}e^{-ik_{1}^{l}x}e^{-i(E+l\hbar
\omega)t/\hbar} \label{8}%
\end{equation}
and%
\begin{equation}
\Psi_{t\text{ }}(x,y,t)=e^{ik_{y}y}%
{\displaystyle\sum\limits_{l=-\infty}^{l=\infty}}
t_{l}\binom{1}{s_{l\text{ }}e^{i\phi_{l}}}e^{ik_{1}^{l}x}e^{-i(E+l\hbar
\omega)t/\hbar} \label{9}%
\end{equation}
where
\begin{align*}
k_{1}^{l}  &  =\sqrt{\left(  \frac{E+l\hbar\omega}{\hbar v_{f}}\right)
^{2}-k_{y}^{2}}\text{ }\\
\phi_{l}  &  =\tan^{-1}(k_{y}/k_{1}^{l})\text{ }\\
s_{l\text{ }}  &  =sgn(E+l\hbar\omega).
\end{align*}
In the barrier region, where $H_{1\text{ \ }}$is nonzero, the eigenfunctions
$\Psi_{b\text{ }}(x,y,t)$ of $H$ can be expressed in terms of the
eigenfunctions $\Psi_{0}(x,y)$ of $H_{0\text{ }}$as\cite{c14}%
\[
\Psi_{b\text{ }}(x,y,t)=\Psi_{0}(x,y)%
{\displaystyle\sum\limits_{n=-\infty}^{n=\infty}}
J_{n}\left(  \frac{V_{1}}{\hbar\omega}\right)  e^{-in\omega t-iEt/\hbar}%
\]
where $J_{n}\left(  \frac{V_{1}}{\hbar\omega}\right)  $ is the $nth$ order
Bessel function. A linear combination of wavefunctions at energies
$E+l\hbar\omega$ $(l=0,\pm1,..)$ has to be taken. Hence

\begin{align}
\Psi_{b\text{ }}(x,y,t)  &  =e^{ik_{y}y}%
{\displaystyle\sum\limits_{l=-\infty}^{l=\infty}}
\left[  B_{l}\binom{1}{s_{l\text{ }}^{\prime}e^{i\phi_{l}^{\prime}}}%
e^{ik_{2}^{l}x}+C_{l}\binom{1}{-s_{l\text{ }}^{\prime}e^{-i^{i\phi_{l}%
^{\prime}}}}e^{-ik_{2}^{l}x}\right] \label{10}\\
&  \times%
{\displaystyle\sum\limits_{n=-\infty}^{n=\infty}}
J_{n}\left(  \frac{V_{1}}{\hbar\omega}\right)  e^{-i(n+l)\omega t-iEt/\hbar
}\nonumber
\end{align}
where
\begin{align*}
\text{ }k_{2}^{l}  &  =\sqrt{\left(  \frac{E-V+l\hbar\omega}{\hbar v_{f}%
}\right)  ^{2}-k_{y}^{2}}\\
\phi_{l}^{\prime}  &  =\tan^{-1}(k_{y}/k_{2}^{l})\\
s_{l\text{ }}^{\prime}  &  =Sgn(E+l\hbar\omega-V).
\end{align*}
The wavefunctions given in equations(\ref{7}-\ref{10}) have to be continuos at
the boundary. Applying this condition at $x=0$ and $x=a$ , i.e. $\Psi_{i\text{
}}(0,y,t)+\Psi_{r\text{ }}(0,y,t)=\Psi_{b\text{ }}(0,y,t)$ and $\Psi_{t\text{
}}(a,y,t)=\Psi_{b\text{ }}(a,y,t)$ and realizing that $\{e^{in\omega t}\}$ are
orthogonal, we obtain the following set of simultaneous equations:%
\begin{align}
A_{n}+r_{n}  &  =%
{\displaystyle\sum\limits_{l=-\infty}^{l=\infty}}
[B_{l}+C_{l}]J_{n-l}\left(  \frac{V_{1}}{\hbar\omega}\right) \label{11}\\
A_{n}e^{i\phi_{n}}-r_{n}e^{-i\phi_{n}}  &  =s_{n\text{ }}%
{\displaystyle\sum\limits_{l=-\infty}^{l=\infty}}
\left[  B_{l}e^{i\phi_{l}^{\prime}}-C_{l}\ e^{-i\phi_{l}^{\prime}}\right]
s_{l\text{ }}^{\prime}J_{n-l}\left(  \frac{V_{1}}{\hbar\omega}\right)
\label{12}\\
\text{here }A_{n}  &  =\delta_{n,0}\nonumber\\
t_{n}e^{ik_{1}^{n}a}  &  =%
{\displaystyle\sum\limits_{l=-\infty}^{l=\infty}}
\left[  B_{l}e^{ik_{2}^{l}a}+C_{l}e^{-ik_{2}^{l}a}\right]  J_{n-l}\left(
\frac{V_{1}}{\hbar\omega}\right) \label{13}\\
t_{n}e^{i\phi_{n}}e^{ik_{1}^{n}a}  &  =s_{n\text{ }}%
{\displaystyle\sum\limits_{l=-\infty}^{l=\infty}}
\left[  B_{l}e^{i\phi_{l}^{\prime}}e^{ik_{2}^{l}a}-C_{l}e^{-i\phi_{l}^{\prime
}}e^{-ik_{2}^{l}a}\right]  s_{l\text{ }}^{\prime}J_{n-l}\left(  \frac{V_{1}%
}{\hbar\omega}\right)  . \label{14}%
\end{align}
The above set has infinite number of coupled equations and contains infinite
number of unknowns( $n,l$ goes from $-\infty$ to $\infty$ ). This linear
system of equations cannot be analytically solved. Nevertheless, the infinite
series in these coupled equations can be truncated and a finite number of
terms starting from $-N$ upto $N$ ,where $N>$ $\frac{V_{1\text{ }}}%
{\hbar\omega}$ , retained if we note that the coupling strength is determined
by the quantity $\frac{V_{1}}{\hbar\omega}$ through Bessel functions
$J_{n}\left(  \frac{V_{1}}{\hbar\omega}\right)  $ and $J_{n}\left(
\frac{V_{1}}{\hbar\omega}\right)  $, they become negligible for order $n$
higher than $V_{1\text{ }}/\hbar\omega.$ Equations(\ref{11}-\ref{14}) are
numerically solved for $t_{n}$. The transmission probability for the $nth$
sideband, $T_{n},$ for which $k_{1}^{n}$ is real and corresponds to
propagating waves, is obtained from:%

\begin{equation}
T_{n}=\frac{\cos\phi_{n}}{\cos\phi_{0}}\left\vert t_{n}\right\vert ^{2}%
\end{equation}
whereas imaginary $k_{1}^{n}$ corresponds to evanescent waves that carry no
particle current with the result $T_{n}=0$. $k_{1}^{n}$ can be real or
imaginary depending on the particular values of the following parameters:
incident energy $E,$ oscillation frequency $\omega,$ incident angle $\phi
_{0}.$ The numerical results obtained are discussed in the next section.
Furthermore, analytical results are obtained if we consider small values of
$\alpha=\frac{V_{1}}{\hbar\omega}\ $and include only the first two sidebands
at energies $E\pm\hbar\omega$ alongwith the central band at energy $E$.
Moreover, we have to invoke the conditions $\hbar\omega<E$ such that
$sgn(E\pm\hbar\omega)=+1$ and $\hbar\omega<\left\vert E-V\right\vert $ such
that $sgn(E-V\pm\hbar\omega)=-1$ for $E<V.$ Hence, we are able to truncate the
sums in equations(\ref{11}-\ref{14}) retaining only the terms corresponding to
the central and first sidebands and obtain analytical results for central and
first sidebands, $t_{0}$ and $t_{\pm1}:$%
\[
t_{0}=\frac{e^{-ik_{1}^{0}a}\cos\theta_{0}\cos\phi_{0}}{\cos\theta_{0}\cos
\phi_{0}\cos[k_{2}^{0}a]+i\sin[k_{2}^{0}a](1+\sin\theta_{0}\sin\phi_{0})}%
\]%
\[
t_{n}=\frac{1}{2}\frac{J_{n}(\alpha)}{J_{0}(\alpha)}\frac{t_{s0}t_{sn}}%
{\cos\phi_{n}}(\Gamma_{n}^{+}+\Gamma_{n}^{-}e^{i(\phi_{0}+\phi_{n})}%
+\Delta_{n}(e^{i\phi_{0}}+e^{i\phi n}))e^{i(\phi_{n}+k_{1}^{0}a)}%
\]
where $n=\pm1$, $t_{s0}$ and $t_{sn}$ are transmission amplitudes for the
static barrier at energy $E$ and $E+n\hbar\omega$ and%
\[
\Gamma_{n}^{\pm}=\Lambda_{n}^{\pm}-\Lambda_{0}^{\pm},
\]%
\[
\Lambda_{n}^{\pm}=\cos[k_{2}^{n}a\pm\theta_{n}]/\cos\theta_{n},
\]%
\[
\Delta_{n}=\Omega_{n}-\Omega_{0},
\]%
\[
\Omega_{n}=i\sin[k_{2}^{n}a]/\cos\theta_{n}.
\]
In the high barrier limit, $\left\vert V\right\vert \gg E$ with the result
$\theta_{0},\theta_{n}\rightarrow0,$ we obtain expressions for transmission
probabilities for the central and the sidebands. For the central band
\begin{equation}
T_{0}\approx\frac{\cos^{2}\phi_{0}}{1-\cos^{2}[k_{2}^{0}a]\sin^{2}\phi_{0}%
}=T_{s0} \label{x}%
\end{equation}
where $T_{s0}$ denotes the transmission probability at incident energy $E$ and
incident angle $\phi_{0}$ in the case of the static barrier. This is the
result obtained as Eq.(4) in\cite{c5}. For sidebands, we obtain:%
\[
\Lambda_{n}^{\pm}=\cos[k_{2}^{n}a]\text{ }\Rightarrow\Gamma_{n}^{\pm}%
=-2\sin[(k_{2}^{n}+k_{2}^{0})a/2]\sin[(k_{2}^{n}-k_{2}^{0})a/2]
\]%
\[
\Omega_{n}=i\sin[k_{2}^{n}a]\text{ }\Rightarrow\Delta_{n}=2i\cos[(k_{2}%
^{n}+k_{2}^{0})a/2]\sin[(k_{2}^{n}-k_{2}^{0})a/2]
\]%
\begin{align*}
t_{n}  &  =\left.  2i\frac{J_{n}(\alpha)}{J_{0}(\alpha)}\frac{t_{s0}t_{sn}%
}{\cos\phi_{n}}\sin[(k_{2}^{n}-k_{2}^{0})a/2](\cos[(k_{2}^{n}+k_{2}%
^{0})a/2]\cos[(\phi_{n}-\phi_{0})/2]\right. \\
&  \left.  +i\sin[(k_{2}^{n}+k_{2}^{0})a/2]\cos[(\phi_{n}+\phi_{0}%
)/2])e^{i(k_{1}^{0}a+(\phi_{0}-\phi_{n})/2)}\right.
\end{align*}
The transmission probability for the sidebands is given by%
\begin{align*}
T_{n}  &  =\frac{\cos(\phi_{n})}{\cos(\phi_{0})}\left\vert t_{n}\right\vert
^{2}\\
&  =\left.  T_{s0}T_{sn}\left(  2\frac{J_{n}(\alpha)}{J_{0}(\alpha)}\right)
^{2}\frac{\sin^{2}[(k_{2}^{n}-k_{2}^{0})a/2]}{\cos\phi_{n}\cos\phi_{0}}%
(\cos^{2}[(k_{2}^{n}+k_{2}^{0})a/2]\cos^{2}[(\phi_{n}-\phi_{0})/2]\right. \\
&  \left.  +\sin^{2}[(k_{2}^{n}+k_{2}^{0})a/2]\cos^{2}[(\phi_{n}+\phi
_{0})/2])\right.
\end{align*}
where $\hbar\omega<E\cos\phi_{0}$ otherwise $T_{-1}=0.$ $T_{sn}=\left\vert
t_{sn}\right\vert ^{2}$ is the transmission probability of electrons at energy
$E+n\hbar\omega$ and incident angle $\phi_{n}$ for the static barrier. We can
also write the above expression as%
\begin{equation}
T_{n}=T_{s0}T_{sn}\left(  2\frac{J_{n}(\alpha)}{J_{0}(\alpha)}\right)
^{2}\frac{\sin^{2}[(k_{2}^{n}-k_{2}^{0})a/2]}{\cos\phi_{n}\cos\phi_{0}}%
(\sin\phi_{0}\sin\phi_{1}\cos^{2}[(k_{2}^{0}+k_{2}^{1})a/2]+\cos^{2}[(\phi
_{0}+\phi_{1})/2]). \label{z}%
\end{equation}
At normal incidence,
\[
T_{\pm1}=\left(  2\frac{J_{\pm1}(\alpha)}{J_{0}(\alpha)}\right)  ^{2}\sin
^{2}[(k_{2}^{0}-k_{2}^{\pm1})a/2]
\]
and if $\hbar\omega<\left\vert E-V\right\vert $ we can write
\[
k_{2}^{0}-k_{2}^{\pm1}=\left\vert E-V\right\vert /\hbar v_{F}-\left\vert
E-V\pm\hbar\omega\right\vert /\hbar v_{F}=\pm\omega/v_{F}%
\]
with the result%
\[
T_{\pm1}=\left(  2\frac{J_{\pm1}(\alpha)}{J_{0}(\alpha)}\right)  ^{2}\sin
^{2}\left[  \frac{\omega a}{2v_{F}}\right]  =\left(  2\frac{J_{\pm1}(\alpha
)}{J_{0}(\alpha)}\right)  ^{2}\sin^{2}[\omega\tau/2]
\]
where $\tau\equiv a/v_{F}$ is the time taken by a normally incident electron
to cross the barrier \textit{without multiple reflections} inside it. From the
above expression, we note that $T_{1}=T_{-1}$. For small $\alpha,$ $J_{\pm
1}(\alpha)\approx\pm\alpha/2;$ $J_{0}(\alpha)\approx1$ and $\sin[\omega
\tau/2]\approx\omega\tau/2$ when $\omega\tau$ is small, corresponding to low
frequency limit where frequency is smaller than the reciprocal of the
traversal time$.$ Using these results we obtain
\[
T_{\pm1}\approx\left(  \frac{V_{1}}{2\hbar}\tau\right)  ^{2}%
\]
The above result can be compared with Eq.(8) in\cite{c16a}, where the
transmission probability through a time-modulated barrier for the first
sidebands is determined. The factor $T$, the transmission probability of the
central band, is not unity and hence it appears there whereas $T_{s0}%
=T_{sn}=1$, for normal incidence, in our case.

\subsection{Results and Discussions}

The results for the transmission of Dirac electrons in graphene are now
presented. The following parameters were used: The Fermi wavelength of the
incident electron is taken to be $\lambda=50nm,$ the barrier oscillation
frequency $\omega=5\times10^{12}Hz,$ the barrier width $a=100nm$ and the
barrier height $V=200meV.$ The dependence of transmission probabilities on
$\alpha=V_{1\text{ }}/\hbar\omega$ for normally incident electrons and for
those arriving at incident angle 30 degrees is shown in Figure(1a,b),
respectively. For normal incidence, the angular dependence of the transmission
probability for the $nth$ sideband is independent of the sign of $n$:
$T_{+n}=T_{-n}$ for $k_{1}^{-|n|}$ real. But this does not hold for incidence
other than normal. We also find that the quantity $\alpha$ is very significant
in determining the relative transmission probabilities of various sidebands as
shown in the figure. This implies that by adjusting the value of $\alpha$ we
can increase transmission through a particular sideband. It is seen that the
central band dominates the transmission at all incident angles for small
values of $\alpha$ whereas contributions from higher and lower sidebands
increases as $\alpha$ becomes larger. This is plausible because for lower
values of $\alpha$ the oscillating barrier can be treated as a static one
since we are keeping $\omega$ fixed and changing $V_{1\text{ }}$ with the
result that $\alpha$ is proportional to $V_{1}$ in these figures. Moreover,
the total transmission probability through the central as well as the
sidebands is unity. Hence, perfect transmission for the oscillating barrier at
normal incidence which was earlier observed for the static barrier\cite{c5}.
This is due to the chiral nature of the particles which results in perfect
transmission (Klein tunneling).

In Figure(2a) we present the angular dependence of the transmission
probability for the central-band $T_{0}$ for various values of $\alpha
=V_{1\text{ }}/\hbar\omega$. The transmission probability for the static
barrier is also shown in the figure as it corresponds to $\alpha=0.$ The
transmission probability $T$ for the static barrier was previously obtained
in\cite{c5}. We find resonant transmission through the oscillating barrier but
unlike the static barrier we do not find perfect transmission for any incident
angle. Realize that for the static barrier there is perfect transmission for
certain values of the incident angle. This is to be expected as the
probabilities are now spread over the central and sidebands. In addition, the
maximum transmission through the oscillating barrier depends on the value of
$\alpha.$

Figure(2b) shows the transmission probabilities for the central band along
with the first few sidebands as a function of the incident angle for
$\alpha=V_{1\text{ }}/\hbar\omega=5.$ In this figure, we show how the incident
particle flux is distributed in the sidebands (through the respective
transmission probabilities) as the incident angle is varied. Note that the
propagation angle for $nth$ sideband is $\phi_{n}$ which is not the same as
the incident angle $\phi_{0\text{ }}.$ For this particular value of $\alpha,$
transmission probability in the central band is very small for normal and
close to normal incidence. For higher sidebands, more and more peaks in
transmission probabilities occur. In the static case, the peaks in the
transmission probability of the central band (there are no sidebands there)
correspond to perfect transmission and the incident angles at which these
occur can be obtained from the resonance condition, $k_{2}^{l}=\frac{p\pi}{a}$
( $p$ is an $\operatorname{integer}),$ through Eq.(\ref{x})and \cite{c5}. For
the time-dependent situation being investigated here, it is not easy to
determine the positions of the peaks as the analytic expression is more
complicated. Nevertheless, we can understand how and where they occur by
examining Eq (\ref{z}), albeit for small $\alpha$ where analytical results can
be obtained but essential physics is the same. We observe, the transmission
probability $T_{n}$ given by Eq.(\ref{z}) depends most strongly on the
prefactor $T_{s0}T_{sn}$ for the parameters considered here. The peaks
correspond to the peak values of $T_{s0}T_{sn}.$ Furthermore, the same
behavior is seen for the static case as the transmission at higher incident
energy there corresponds to transmission in the sidebands here. At these
higher energies, the $x$-component of momentum in the barrier region satisfies
the resonance condition greater number of times as the incident angle is
varied, thus larger number of peaks.

We note that the absence of any potential gradient along the $y$-direction
results in the conservation of the $y$-component of momentum. Therefore,
change in energy that an electron experiences due to exchange of modulation
quanta with the oscillating barrier brings about corresponding changes only in
the $x$-component of the electron's momentum. For non-zero $k_{y}$, energy
exchanges can makes $x$-component of momentum imaginary inside or/and outside
the barrier region that corresponds to unavailability of any energy state in
the relevant region(s). If energy $E+l\hbar\omega$ in the $lth$ sideband is
such that $\left\vert E+l\hbar\omega\right\vert <\hbar v_{F}k_{y},$ there are
no propagating states available outside the barrier since $k_{1}^{l}$ becomes
imaginary. At the same energy when particles have states available inside the
barrier it can be localized if it is transferred to these states after losing
energy through interaction with the oscillating barrier. In this situation,
the particles are confined across the barrier while they are free to propagate
along the barrier till one or more quantum of energy is absorbed, allowing
transition to a higher sideband with states aligned in energy outside the
barrier leading to eventual escape from the barrier region. For a graphene
quantum well, confined electron states which arise due to the suppression of
electron-hole conversion at the barrier have been discussed in \cite{c24}.

For electron energy such that $\left\vert E+l\hbar\omega-V\right\vert <\hbar
v_{F}k_{y},$ there are no propagating states available inside the barrier
since $k_{2}^{l}$ becomes imaginary. Furthermore, the energy at which
electronic states outside the barrier match the hole states inside it,
electronic transmission is governed by Klein tunneling while unavailability of
hole states inside the barrier results in ordinary tunneling.

In Figure (3a) we present the transmission probability as a function of
barrier width $a$ for normal incidence. For the static barrier there is
perfect transmission as can be seen in Figure (3a) where $T$ represents the
transmission probability for the static barrier whereas the transmission
probability for the central band in the oscillating barrier decreases for
smaller values of the barrier width and shows oscillatory but damped behavior
for larger barrier width. The transmission probability for the other sidebands
increases initially from zero but then oscillates with damped amplitude. We
also observe that the contribution in transmission of the higher sidebands
rises as the barrier width increases, this occurs due to larger time available
to the electron for interacting with the oscillating potential as it traverses
the barrier. In addition, we find that for normal incidence in the oscillating
barrier: $T_{+n}=T_{-n}$ for $k_{1}^{-|n|}$ real. Nevertheless, the total
transmission probability through the central as well as the sidebands is
unity. These results imply that perfect transmission at normal incidence is
independent of the barrier width, yet another manifestation of Klein tunneling.

In Figure(3b), the transmission probability as a function of barrier width $a$
when the incident angle is $30$ degrees is shown. The transmission probability
represented by $T$ for the static barrier now oscillates as a function of the
barrier width whereas transmission probabilities for the central and sidebands
in the oscillating barrier show behavior close to that obtained for normal incidence.

A comparison between analytical result obtained in Eq(\ref{z}) and numerical
results is presented in Figure(4) for $\alpha=0.5$. Transmission probabilities
$T_{\pm1}$ of first sidebands are plotted against incident angle $\phi_{0}$.
Inset shows plot of $T_{-1}$ versus $\phi_{0}.$ It shows that transmission
probabilities determined numerically exhibit the same behavior as obtained in
the analytical result.

To summarize, we have considered the tunneling of chiral massless electrons
corresponding to monolayer graphene through a barrier that is oscillating
harmonically in time. We have determined how the transmission probability for
the central and sidebands depends on the incident angle of the particles, the
width of the barrier, the height and frequency with which it oscillates. Due
to the chiral nature of the particles in graphene, tunneling is highly
anisotropic with peculiar behavior at normal and close to normal
incidence(Klein tunneling). We find, for normal incidence, perfect
transmission in monolayer graphene. Klein tunneling that was observed for the
static barrier is found to persist for the oscillating barrier.

\section{Acknowledgements}

One of us (K.S.) would like to acknowledge the support of the Pakistan Science
Foundation (PSF) through project No. C-QU/Phys (129).

$\dagger$corresponding author: ksabeeh@qau.edu.pk, kashifsabeeh@hotmail.com.

\end{document}